\def\mincir{\ \raise -2.truept\hbox{\rlap{\hbox{$\sim$}}\raise5.truept	
\hbox{$<$}\ }}								%
\def\magcir{\ \raise -2.truept\hbox{\rlap{\hbox{$\sim$}}\raise5.truept	%
\hbox{$>$}\ }}								%

\documentstyle[11pt,newpasp,twoside,epsf]{article}
\def\edcomment#1{\iffalse\marginpar{\raggedright\sl#1\/}\else\relax\fi}
\reversemarginpar

\begin{document}
\title{The Connection between BL Lacs and FSRQs}
 \author{V. D'Elia, A. Cavaliere}
\affil{Astrofisica, Dip. Fisica, Universit\'a Tor Vergata, Roma, I-00133}

\begin{abstract}
We discuss the features that mark the Flat Spectrum Radio Quasars 
from the BL Lacertae
objects. We propose that FSRQs exceeding 
$L \sim 10^{46}$ erg s$^{-1}$ are powered by black hole 
accreting at rates $\dot m \sim 1 \div 10$; their 
power is dominated by the disk components, 
thermal and BZ. 
Instead, sources accreting at rates $\dot m \sim 10^{-2} \div 10^{-3}$ radiate 
in the BL Lac mode; 
here the power is mainly non-thermal, and is driven 
by the rotational energy stored in a Kerr hole which sustains 
 $L \mincir 10^{46}$ erg s$^{-1}$ in the jet frame for Gyrs.
The two populations may be even linked if around the same 
objects $\dot m$ drops in time; then 
some negative cosmological evolution 
is expected for the newborn BL Lacs fed by 
the dying FSRQs. Further implications are discussed. 

      
\end{abstract}

\section{Introduction}

Within the AGNs, the Blazars are singled out by their 
flat-spectrum GHz emission, powerful
$\gamma$ rays into the GeV band and beyond, rapid variability, 
high and variable optical polarization. 

These common properties are widely explained (see Urry \& Padovani 1995) 
in terms of beamed emissions from a relativistic jet, with Lorentz bulk
factors $\Gamma \approx 5 \div 20$ (in the following, we will
concentrate on the debeamed e.m. power). 

Within the Blazars, the Flat Spectrum Radioloud Quasars (FSRQs) differ from the 
BL Lac objects on the following accounts:
1) optical features
2) integrated power
3) top spectral energies
4) cosmological evolution.     
The specific features of the two classes are collected in the following
Table 1:

\medskip 
\centerline{\bf Table 1 - Observational features of FSRQs and BL Lacs}
\begin{table}[h]
\hspace{1.0cm} 
\begin{tabular}{|l|c||c|}
\hline
          &FSRQs  &BL Lacs  \\
\hline
\hline
optical features      &em. lines, bump & no lines, no bump\\
\hline
integrated power      &$L \sim 10^{46 \div 48}$ erg s$^{-1}$
		&$L \mincir 10^{46}$ erg s$^{-1}$  \\
			
\hline
evolution             & strong & weak if any \\		
\hline
top energies     & $h\nu \sim  10$ GeV & $h\nu \sim 10 $ TeV\\  

\hline

\end{tabular}
\end{table}

These differences stand out of selections and call for explanation. 
Within the accreting black hole paradigm for the primary power source, 
we shall order all these features in a {\it sequence} primarily marked by 
one parameter, the accretion rate $\dot m$ (Eddington units). We 
propose that the FSRQs, like other quasars, accrete at rates $\dot m \sim 1 
\div 10$, 
while the BL Lacs radiate in conditions of $\dot m \ll 1$. 
 
\section{The Blazar Luminosities}

We begin with the {\it thermal} luminosities, including the optical-UV bumps; 
in the standard view these are produced in the accretion {\it disk} and are 
given by $L_{th} \approx \dot m \,L_{E}$. 
Then the absence or weakness of bumps in the  
BL Lac spectra can be simply understood in terms of $\dot m \ll 1$. 
If so, low gas densities 
are also expected around the hole, and these concur to account for the other 
optical feature of 
the BL Lacs, namely, the weak or absent emission lines. The
FSRQs instead, in spite of their 
``blazing'' non-thermal component,
share with other quasars the bumps and the broad emission 
lines; these features are consistent with values $\dot m \sim 1 \div 10$.

As to the {\it jets}, we state our guideline: jets are powered by 
variants of the 
mechanism  originally proposed by Blandford \& Znajek (1977) for 
extraction of rotational energy from a Kerr hole via the Poynting-like flux 
associated with the 
surrounding magnetosphere. 
Variants are necessary in view of the limitations recently discussed 
to the power extractable from the hole. Since the BZ power scales as  
$ B_h^2 (r_c/r_h)^2$, such variants involve either high strengths  
of the magnetic fields $B_h$ threading the hole horizon at $r_h$, or the MHD 
contribution of the disk from a radius $r_c$ larger than $r_h$. 

Recent discussions (Modersky \& Sikora
1996; Ghosh \& Abramowicz 1997; Livio, Ogilvie \& Pringle 1999) have stressed  
the continuity of $B_h$ with the field $B_d$ rooted in the inner stable region of 
the disk; in turn, $B_d$ is bounded after 
$B_d/8\pi \mincir P_{max}$ by the maximum pressure. 
In a standard $\alpha$-disk the latter scales as 
$P_{max} \propto (\alpha M_9)^{-9/10}{\dot m}_{-4}^{4/5}$
if it is gas dominated, or as
$P_{max} \propto (\alpha M_9)^{-1}$ if it is radiation dominated; 
in the latter case the {\it hole} power attains its maximum 
$$ L_{K}=2 \, 10^{45}\,M_9 \,(J/J_{max})^2\,  erg s^{-1} ~.\eqno(1)$$  

But to have an inner disk region dominated by radiation pressure, 
accretion rates $\dot m \magcir 10^{-3}$ are required. This is because
the radius $r_c$ bounding the region
grows with $\dot m^{16/21}$ (Novikov \& Thorne 
1973), and exceeds the last stable orbit only if
$\dot m \magcir 10^{-3}(\alpha M_9)^{-1/8}$ holds.
In such conditions, the hole output can exceed the disk thermal luminosity as 
given by 
$L_K/L_{th} = 3.3\,10^{-2} \dot m^{-1}\,(J/J_{max})^2.$

With $\dot m $ increasing, $L_K$ saturates and its 
ratio to $L_{th}$ decreases, but the radiation pressure region tends to 
broaden. Then 
a larger power component may be extracted from the disk, up to 
$L_d \sim L_K \, (r_c/r_h)^2$. 

Within this framework, we draw the following 
implications concerning the Blazar sequence. BL Lac jets can live 
with accretion rates $\dot m \sim 10^{-2}$, since  
the rotational power levels $L_K$ given by eq. (1) are 
often adequate even considering the kinetic power remaining in the jets 
(see Celotti 1999). The most powerful BL Lacs 
require very massive BHs and/or 
a larger but still comparable contribution from the dynamically 
 entrained and magnetically connected inner rings of the disk. In all such 
cases one expects  
$L_d \sim L_{K} \magcir L_{th}$ to hold.

On the other hand, many FSRQs feature 
outputs exceeding $10^{46}$ erg s$^{-1}$ considerably, 
with specific sources approaching a total of $10^{48}$ erg s$^{-1}$
(Tavecchio et al. 2000).  
Such outputs require  dominant components 
$L_d \sim L_{th}\gg L_K$ from a 
wider disk region dominated by radiation pressure, 
and so require conditions where $\dot m \sim 1 \div 10$.
The hole contribution, though minor, is likely to 
provide a ``high-velocity spine'' 
instrumental for the jet propagation, see Livio 1999. 

Alternatively, to account for such huge outputs one needs
very strong $B_h$, as advocated by Meier 1999; 
fields up to $B_h^2/8\pi\sim \rho c^2$ 
in the plunging 
orbit region have been argued by Krolik 1999; 
Armitage, Reynolds \& Chiang 2000 and Paczynski 2000 discuss 
how and why such enhancements are unlikely inside  
a thin disk. In thick disks the status of such enhanced fields is still 
uncertain; we note they would require high $\dot m$ anyway.

In this section we have shown how different values of the key parameter 
$\dot m$ mark  
thermal and non-thermal features {\it together} 
along the Blazar sequence. We summarize our discussion 
by adding to the previous Table 1 the following lines:

\begin{table}[h]
\hspace{2.0cm} 
\begin{tabular}{|l|c|c|c|}
\hline
          &FSRQs  &BL Lacs \\
\hline
Kerr hole vs. disk           &$L_K \ll L_d$     & $L_{K} \mincir L_d$ \\
\hline
key parameter: $\dot m$      &$\dot m \sim 1 \div 10$      &$\dot m \sim 10^{-3} 
\div 10^{-2} $ \\
\hline
\end{tabular}
\end{table}

\section{The Blazar Evolutions}

Strong cosmological 
evolution is closely shared by the FSRQs with the rest of the quasars 
(Goldschmidt et al. 1999), 
and shows up, e.g., in their steep number counts;
the BL Lacs, instead, 
show no signs of a similar behavior (Giommi, Menna \& Padovani 1999; 
Padovani 2000). 

The quasar behavior includes a strong component of luminosity evolution
(see Boyle 
et al. 2000 for the optical and Della Ceca et al. 1994 for the X-ray band). 
This is widely traced back to the exhaustion in the host galaxy 
of the gas stockpile usable for accretion, due to previous accretion episodes 
and to star formation (Cattaneo, Haehnelt \& Rees 1999; 
Haehnelt \& Kauffmann 2000; Cavaliere {\&} Vittorini 2000). 
With the average rate $\dot m$ so decreasing, we expect many objects 
to switch from being mostly fueled by 
accretion to being mostly fed by the Kerr hole rotational supply 
$E_K$ (stored 
by accretion of angular momentum $J$ along with mass, Bardeen 1970); so
we expect many sources to switch from the FSRQ to the BL Lac mode. 
The moderate BL Lac powers  
can be sustained for several Gyrs by the coupled system 
Kerr hole - disk, so the BL Lac luminosity evolution 
is expected to be slow (Cavaliere {\&} Malquori 1999), 
with time scales around $\tau_L \mincir E_{K}/L_K \approx 8$ Gyr.  

In more detail, Cavaliere \& Vittorini  2000 
trace back the bright quasar evolution to the diminishing
rate of the interaction episodes of the host galaxies with their 
neighbors in a group. These events destabilize the host gas and trigger 
accretion; they last some $10^{-1}$ Gyr, a galactic dynamical 
time, and produce a weak {\it density} evolution on a time scale 
 $\tau_D \approx 6$ Gyr. But the efficiency of such episodes  
drops, due to the exaustion of the
gas available in the hosts over times $\tau_{L}\approx 3$ Gyr; this 
produces a strong {\it luminosity} evolution. 

Our point is that the powerful FSRQ activity based on high $\dot m$ 
will die out over times of some $10^{-1}$ Gyr after a ``last interaction'';
but in many instances this will leave behind a maximally 
spinning hole, and so
a long lived BL Lac. Thus the scale $\tau_D$ for bright FSRQ deaths 
is also the scale for BL Lac births.  

One sign of evolution is provided by the integrated counts, which 
at high/ medium fluxes may be evaluated from the crude but explicit expression 
$$ N(>S) \propto 
S^{-3/2}[1-C(S_0/S)^{1/2} + 0(S^{-1})]~,  \eqno (2)$$ 
with the key time scales directly appearing in the coefficient 
$$ C = 3\,D_0\langle l^2 \rangle [2(1+\alpha) -  1 /H_0\, \tau_D 
-  (\beta-1) / H_0\, \tau_L]/4\,R_H\langle l^{3/2} \rangle~.  
\eqno(3)$$

For the BL Lacs counted in the radio band 
such time scales are: 
$\tau_D \approx -\, 6$ Gyr (BL Lac births imply 
{\it negative} density evolution); $\tau_L\approx 8$ Gyr 
(marking the slow BL Lac luminosity evolution).
Other quantities involved are: the 
spectral index $\alpha=0.3$ in the GHz range; 
the slope $\beta \approx 2.5$ of the radio LF;
the normalized moments $\langle l^n\rangle$ of the LF; the distance $D_0 = 
(L_0/4 \pi S_0)^{1/2}  \approx 0.05 R_H$ 
in Hubble units of typical high flux BL Lacs.  
 
For example, in the critical universe with $t_0 \approx 13$ Gyr
the result is $C\approx 0.1$.
This means $N(>S) \propto S^{-1.5}$ or flatter at high fluxes,  
consistent with the data by Giommi et al. 1999, see fig. 1. 

In contrast, the values appropriate for the FSRQs, namely:  
 $\tau_D = +\, 6$ Gyr, $\tau_L = 3$ Gyr, 
$ D_0 \approx 0.5 R_H$  
and still $\beta = 2.5$, 
yield $C \approx 4 $ and produce radio counts $N(>S)$ steepening well above 
 $ S^{-1.5}$. 

Note that $C$ includes $\beta$ and 
$\langle l^2 \rangle/\langle l^{3/2} \rangle$, 
which both act to
flatten the counts for flatter LFs. 
In fact, the beaming effect 
(Urry {\&} Padovani 1995) does flatten the LF at the faint end, which 
contributes to the flattening of the faint counts; 
however, the similarly affected FSRQ counts show a bright steep 
section indicative of intrinsically stronger evolution. 

To summarize this discussion, we add to Table 1 the following line:

\begin{table}[h]
\hspace{3.0cm} 
\begin{tabular}{|l|c|c|c|}
\hline
          &FSRQs  &BL Lacs \\
\hline
evolution      &strong
		&weak if any \\
\hline
\end{tabular}
\end{table}

\section{The Blazar Spectra: $\gamma$ rays}

Another feature 
of the Blazars is their SED that extends into the 10 GeV 
range for the FSRQs, and into the TeV range for the BL Lac objects. 
Such high energy photons are likely produced via inverse Compton 
(Ghisellini 1999) by GeV 
and by $10^2$ GeV electrons, respectively.
Can the parameter $\dot m$ also explain this difference? 

To such energies the particles may be accelerated in two ways: 
either by weak electric fields ($E \sim 10^{-8}$ 
cgs units) over large distances ($ \sim 10^{16}$ cm) as in the internal shock 
scenario which, however, falls short of the 
top energies required in BL Lacs (Ghisellini 1999);
or by higher fields (associated with
energy transport via ``Pointing flux'' along the jet) effective 
over shorter distances. 

In pursuing the latter way, we expect the force-free condition 
$ E \bullet B = 0$ governing the BZ magnetosphere 
to break down at average distances $R \sim 10^{17}$ cm, but to do it
inhomogeneously within bubbles or filaments; 
however, fields with natural values $E \sim 1 $ cgs units 
still would be screened out over distances exceeding some
$c/\omega_p \propto (\gamma/n)^{1/2}$. The densities 
may be estimated from the emissions which scale
as $L\sim \gamma^2 \, U\,  R^3\, n $ with $\gamma^2UR^3$ roughly
constant (Fossati et al. 1999; Ghisellini 1999). So in comparing 
BL Lacs as a  class with the FSRQs, values of $n$ smaller by $10^{-3}$ obtain;
then the top electron (and $\gamma$-ray) energies ought to 
scale like ${\cal E}_{max} \propto (\gamma^2UR^3)^{1/2}\;\gamma_{min}^{1/2}
\;L^{-1/2}$.  
The scatter of $\nu_{peak}$ in the BL Lac class expected from the second 
parameter $L_d/L_K$ will be discussed elsewhere.

After this discussion we may 
add to Table 1 the line:

\begin{table}[h]
\hspace{2.5cm} 
\begin{tabular}{|l|c|c|c|}
\hline
          &FSRQs  &BL Lacs \\
\hline
top energies      &$h\nu \sim  10$ GeV  &$h\nu \sim 10$ TeV \\ 
		
\hline
\end{tabular}
\end{table}

\begin{figure}
\epsfysize=7cm 
\hspace{3.5cm}\epsfbox{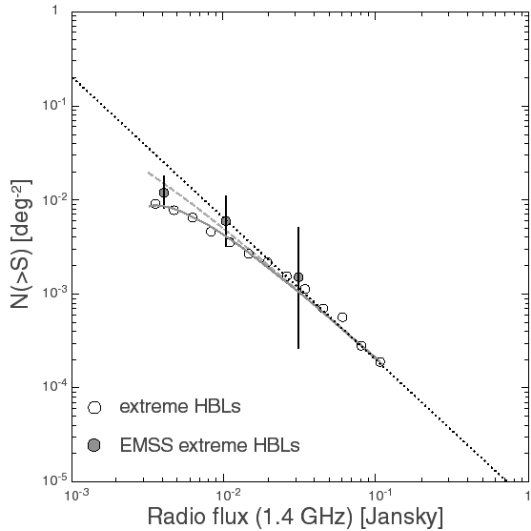} 
\caption[h]{The BL Lac counts evaluated from eqs. (2) and (3) 
using $\tau_D = - \, 7$ Gyr (dashed)
and $\tau_D = -\, 5 $ Gyr (solid). The dotted line represents the 
``Euclidean'' slope. Data for extreme BL Lacs from Giommi, Menna, {\&} 
Padovani 1999.}
\end{figure}  

\section{A Link with Accelerators of UHECR?}

If we carry the Blazar sequence to its extreme, we may expect endpoint objects 
(dead BL Lacs) with
very low residual $\dot m \mincir 10^{-4} $, 
which would feature very faint if any e.m. emission,
along with nearly unscreened electric fields. These would accelerate particles 
including protons, and make ultra high energy cosmic rays up 
to the long recognized limit 
given by ${\cal E}_{max} \sim e\,B_D\,r_{ms}\,(r_{ms}/R)^p/p \sim 10^{20}\,M_8\,B_4$
eV; after Blandford {\&} Payne 1982, the magnetic fields are assumed to 
decrease outwards like $r^{-(1+p)}$ with $p=1/4$. 

Several tens of these accelerators could lie within
some 100 Mpc (Boldt {\&} Ghosh 1999). As to ultra high energies,  
these accelerators would evade in the simplest way 
the GZK cutoff (if it exists); as to accounting for the particle 
flux, they need to produce only  
$L\sim 10^{42}$ erg s$^{-1}$. Note that nG intergalatic fields 
would blur the geometrical memory of the sources for most but not all UHECRs. 

\section{Conclusions}

The lines added to Table 1 as a summary of each of the Sects.  2, 3, 4,  
5 lead us to propose the overall Table 2:

\medskip
\centerline{\bf Table 2 - The Blazar sequence }
\begin{table}[h]
\hspace{1.0cm} 
\begin{tabular}{|l|c|c|c|}
\hline
          &FSRQs  &BL Lacs & $\rightarrow$ CR accelerators \\
\hline
optical features      &em. lines, bump & no lines, no bump   &  none    \\
\hline
integrated power      &$L \sim 10^{47 \div 48}$ erg s$^{-1}$
		&$L \mincir 10^{46}$ erg s$^{-1}$ &
		$L \mincir 10^{42}$ erg s$^{-1}$ \\
\hline
evolution             & strong &weak if any &negligible \\		
\hline
top energies              & $h\nu \sim 10$  GeV   & $h\nu \sim 10$ TeV  
			&  ${\cal E}_{max}\sim 10^{21}$ eV  \\
\hline
Kerr hole vs. disk           &$L_K \ll L_D$     & $L_{K} \mincir L_D$ 
                           &very low $L_{K}$ and $L_D$ \\
\hline
key parameter: $\dot m$      &$\dot m \sim 1 \div 10$      &$\dot m \sim 10^{-2}$
   &$\dot m \mincir 10^{-4}$     \\
\hline
\end{tabular}
\end{table}
\noindent
We thank P. Giommi for discussions of his data
and for critical reading, and M. Salvati for helpful exchanges. Grants 
from ASI and MURST are acknowledged.

\end{document}